\documentclass[a4paper,man,natbib,floatsintext]{apa6}
\usepackage{apacite}
\usepackage{subfig}
\usepackage[english]{babel}
\usepackage[utf8x]{inputenc}
\usepackage{amsmath}
\usepackage{graphicx}
\usepackage[colorinlistoftodos]{todonotes}
\usepackage[left]{lineno}
\usepackage{diagbox}
\usepackage{indentfirst}
\usepackage{adjustbox}
\usepackage{xcolor}
\usepackage{multirow}

\usepackage{setspace}

\begin{document}

\raggedbottom

 \captionsetup[table]{
  labelsep = newline,
  textfont = sc, 
  name = TABLE, 
  justification=raggedleft,
  singlelinecheck=off,
  labelsep=colon,
  skip = \medskipamount}

\begin{titlepage}

\begin{center}

\large Psychophysiological responses to takeover requests in conditionally automated driving
\\ 

\normalsize

\vspace{25pt}
Na Du\\
Industrial and Operations Engineering, University of Michigan\\
\vspace{15pt}
X. Jessie Yang\\
Industrial and Operations Engineering, University of Michigan\\
\vspace{15pt}
Feng Zhou\\
Industrial and Manufacturing Systems Engineering, University of Michigan-Dearborn\\
\vspace{15pt}

\end{center}
\begin{flushleft}
\textbf{\textit{Accepted to be published in Accident Analysis \& Prevention 09/23/2020}} \\
\textbf{Manuscript type:} \textit{Research Article}\\
\textbf{Running head:} \textit{Psychophysiological responses to takeover requests}\\
\textbf{Word count:} 5800 \\

\textbf{Corresponding author:} 
Feng Zhou, 4901 Evergreen Road, Dearborn, MI 48128, Email: fezhou@umich.edu

\textbf{Acknowledgment:} This work was supported by University of Michigan Mcity.

\end{flushleft}

\end{titlepage}
\shorttitle{}

\section{ABSTRACT}

In SAE Level 3 automated driving, taking over control from automation raises significant safety concerns because drivers out of the vehicle control loop 
have difficulty negotiating takeover transitions. 
Existing studies on takeover transitions have focused on drivers' behavioral responses to takeover requests (TORs). As a complement, this exploratory study aimed to examine drivers' psychophysiological responses to TORs as a result of varying non-driving-related tasks (NDRTs), traffic density and TOR lead time. A total number of 102 drivers were recruited and each of them experienced 8 takeover events in a high fidelity fixed-base driving simulator. Drivers' gaze behaviors, heart rate (HR) activities, galvanic skin responses (GSRs), and facial expressions were recorded and analyzed during two stages.
First, during the automated driving stage, we found that drivers had lower heart rate variability, narrower horizontal gaze dispersion, and shorter eyes-on-road time when they had a high level of cognitive load relative to a low level of cognitive load. Second, during the takeover transition stage, 4s lead time led to inhibited blink numbers and larger maximum and mean GSR phasic activation compared to 7s lead time, whilst heavy traffic density resulted in increased HR acceleration patterns than light traffic density.
Our results showed that psychophysiological measures can indicate specific internal states of drivers, including their workload, emotions, attention, and situation awareness in a continuous, non-invasive and real-time manner. 
The findings provide additional support for the value of using psychophysiological measures in automated driving and for future applications in driver monitoring systems and adaptive alert systems.





\textbf{Keywords:} Human-automation interaction, Automated driving, Transition of control, Psychophysiological measures.


\newpage

\section{1. INTRODUCTION}
The introduction of automated features in vehicles represents a new era for the automotive industry. While we are still a long way off from fully automated vehicles, vehicles with SAE Level 3 automation, such as the Audi A8 Traffic Jam Pilot, have been developed. 
They allow drivers to move their eyes from the road and hands off the steering wheel \citep{sae}. However, such SAE Level 3 automated vehicles as Audi A8 with all the technology available to make Traffic Jam Pilot work, have not been on the road for usage due to challenges during takeover transitions \citep{pete2019}. 


In conditionally automated driving, when the driver is out of the vehicle control loop, s/he lacks sufficient situation awareness of the driving environment. Once the vehicle reaches the operational limit of the automated driving system, the vehicle will request the driver to take over control from the automated driving. Under such circumstances, the driver often has difficulty negotiating the takeover transitions safely \citep{Ayoub:2019:MDA:3342197.3344529,janssen2019interrupted,seppelt2019keeping, Zhou2020}. 
To evaluate drivers' takeover performance, existing literature has measured various types of driving behaviors such as takeover reaction time, maximum resulting acceleration, and minimum time to collision \citep{du2020examining,gold2016taking,clark2017age,wan2018effects,naujoks2014effect}. 

While driving behaviors alone shed light on drivers' takeover performance, psychophysiological measures have their sensitivity and specificity to provide us a broad picture of the internal states (e.g., cognitive workload, emotions, attention, and situational awareness) that drivers experience. 
This exploratory study aimed to examine the effects of non-driving-related tasks (NDRTs), traffic density, and takeover request (TOR) lead time on drivers' psychophysiological responses to TORs in simulated SAE Level 3 automated driving. The inclusion of psychophysiological measures can complement takeover performance measures and help us understand drivers' state-level changes timely and continuously. 

The rest of this paper is organized as follows. The remaining part of Section 1 gives the background for the work and an overview of the present study. Section 2 describes the method, including experiment design and data analysis. The results are presented in Section 3 and are discussed in Section 4. We conclude the paper in Section 5.

\subsection{1.1. Takeover performance measurements}
Drivers' takeover transitions in conditionally automated driving can be affected by many factors, including drivers' characteristics (e.g., age, gender), types of NDRTs (e.g., cognitive load and emotional states triggered by NDRTs), vehicle configuration (e.g., TOR lead time, TOR modality), and driving environments (e.g., traffic density, weather) \citep{wu2020age,du2020examining,gold2016taking,li2018investigation}. To quantify how these factors influence takeover transitions, existing studies have mainly focused on driving behaviors after TORs. 
Driving behaviors are categorized into two aspects, namely, takeover timeliness and takeover quality for takeover performance measurements. Takeover timeliness means how quickly drivers respond to TORs and is measured as the time between the TOR and the first indicator of takeover maneuver. Takeover quality consists of a wide range of metrics including speed, acceleration and jerk statistics, time/distance to collision statistics, steering angle and pedal statistics, lane deviation statistics, and crash rate. For example,  
\cite{gold2016taking} measured drivers' minimum time to collision (TTC) and crash numbers and illustrated that heavy traffic density led to worse takeover quality demonstrated by shorter minimum TTC and more crashes. More recently, \cite{du2020examining} used smaller maximum resulting acceleration and maximum resulting jerk as indicators of good takeover quality to show the advantages of positive emotional valence for takeovers during automated driving. 

While these driving metrics quantify drivers' vehicle control after TORs and provide insight into the prominent effects of factors on takeover performance, they have the following limitations. First, driving metrics capture drivers' behaviors at the specific moment (e.g., minimum TTC) or at the overall level (e.g., standard deviation of lane positions), but lack understanding of the entire takeover process in a consecutive time-series way. Second, although drivers sometimes do not show observable varieties at the performance level, their cognitive and emotional states might be significantly influenced and should be used to measure their overall takeover experience. 
Self-reported subjective measures can also assess drivers' internal states. Yet, self-reporting internal states significantly interferes with the real-time task at hand and could be difficult for drivers during the takeover transitions \citep{schmidt2009drivers}. Therefore, it is necessary to collect drivers' psychophysiological signals to examine their workload, emotions, attention, and situation awareness, timely and continuously. 



\subsection{1.2. Psychophysiological measurements in driving}

With the development of low-cost and non-invasive wearable sensors, it is achievable to collect drivers' psychophysiological signals to reflect their states affected by NDRTs, vehicle configurations, and driving environments. Commonly used measurements in vehicle-related research include eye movements, heart rate (HR) activities, galvanic skin responses (GSRs), facial expressions, and so forth. 

Gaze behaviors, such as gaze dispersion and blink number, have been widely used in driving studies to reflect drivers' cognitive load, attention, and situational awareness \citep{wang2014sensitivity,luo2019toward,young2013missing,lemercier2014inattention}. Researchers have shown that increases in drivers’ cognitive load induced by NDRTs and environments are linked to increases in pupil diameter and decreases in horizontal gaze dispersion and blink number \citep{wang2014sensitivity,gold2016taking,merat2012highly,luo2019toward}. For example, \cite{merat2012highly} compared drivers' states when they were in different scenarios (with vs. without critical incident), NDRTs (with vs. without Twenty Questions Task), and drive (manual vs. automated). They found that blink frequency was generally suppressed during high workload conditions, where drivers experienced critical incidents and Twenty Questions Task.
Regarding the attention perspective, \cite{Louw2015} investigated driver attention in automated driving and measured drivers' gaze dispersion with four manipulations: 1) no manipulation, 2) light fog, 3) heavy fog, and 4) heavy fog with a visual NDRT. They found that drivers had wider gaze dispersion when the driving scene was completely in the heavy fog condition, but became more concentrated if a visual NDRT existed. Although gaze dispersion and eyes-on-road time percentage are traditionally treated as distraction indicators in manual driving, wider gaze dispersion and larger eyes-on-road time percentage imply high situation awareness in automated driving \citep{young2013missing,molnar2017age}.

Heart rate and heart rate variability (HRV) have the sensitivity to assess drivers' workload and detect workload changes before the presence of observable effects in driving performance \citep{mehler2012sensitivity, mehler2009impact,bashiri2014heart,lohani2019review,hidalgo2019respiration}.
For instance, \cite{hidalgo2019respiration} conducted a driving simulator study with 18 subjects and found that decreases in HRV were associated with increases of cognitive load during manual driving. More importantly, HRV reflected such variations in attention and cognitive load levels before differences in driving performance was evident.
Although some researchers have argued that cardiac responses remain open for attention interpretation, it is widely established that heart rate acceleration and deceleration are associated with defense and orienting responses, respectively. 
Specifically, Lacey and Sokolov proposed that heart rate acceleration occurred in situations involving stimulus ignorance and environmental rejection, while heart rate deceleration indexed the intake and enhancement of environmental stimuli \citep{libby1973pupillary,lacey1970some,sokolov1963higher,sokolov1961progressive}. Take the driving context for an example, \cite{reimer2011impact} 
found that younger drivers had heart rate acceleration in response to the phone conversation task in simulated manual driving. This pattern indicated that drivers selectively ignored or rejected disruptive input, which was the phone task in this setting. However, late middle aged drivers did not demonstrate such a pattern possibly due to individual differences in attentional focuses.

Galvanic skin responses (GSRs) measure skin conductance controlled by changes in the sympathetic nervous system. Raw GSR signals comprise of two components, i.e., phasic activation (rapid changes to a specific stimulus) and tonic activation (slower responses at background level of the activity) \citep{boucsein2012electrodermal}. 
GSRs have been found to be associated with drivers’ cognitive load, stress, and emotional arousal \citep{collet2009physiological,mehler2012sensitivity,wintersberger2018let}. For example, \cite{mehler2012sensitivity} conducted an on-road study where 108 drivers across three age groups performed an auditory working memory task with three difficulty levels during manual driving. Results showed that drivers had increased heart rate and skin conductance with a high level of cognitive demand. In the context of automated driving, \cite{wintersberger2018let} measured drivers' GSRs after TORs in a simulated driving study. They found that GSR phasic activation, as an indicator of drivers' arousal and stress, became higher when TORs were presented during an NDRT than between NDRTs.


Facial expressions have been used to recognize drivers' and passengers' emotional states in driving \citep{wintersberger2016automated,gao2014detecting,izquierdo2018emotion}. For example, \cite{wintersberger2016automated} made use of passengers'  facial expressions to estimate their emotional responses (in pleasure and arousal dimensions) when they were in a vehicle driven by an automated driving system, a male, or a female driver. Furthermore, \cite{izquierdo2018emotion} developed a k-Nearest Neighbors algorithm to classify drivers' emotions (e.g., anger, sad, joy, anxiety) in automated driving using facial expressions and reached an accuracy of approximately 97\%. Such models can potentially be used to understand drivers’ emotional states and the vehicle might respond in real time to improve drivers' user experience and reduce possible aggressive behaviors (e.g., when in agner).



\subsection{1.3. The present study}
Existing studies on drivers’ responses to TORs mainly focused on their takeover performance. Little is known about drivers’ cognitive load, attention styles, and emotional states amid takeover transitions, which  can be reflected through psychophysiological measurements though. In addition, those studies that reported psychophysiological signals in driving mostly focused on manual driving and did not show the psychophysiological results in a systematic and time-series manner \citep{hidalgo2019respiration,reimer2011impact,mehler2012sensitivity}.



This exploratory study aimed to examine drivers' psychophysiological responses to TORs in different NDRTs, traffic density, and TOR lead time conditions. Psychophysiological data collected in the study included drivers' gaze behaviors, HR activities, GSRs, and facial expressions. A total number of 102 drivers participated in the study and each experienced eight takeover scenarios in a high fidelity driving simulator. Before takeover performance showed observable discrimination, psychophysiological signals collected by non-intrusive sensors showed the advantages to enable continuous and real-time assessment of drivers' cognitive workload, emotions, attention, and situational awareness during the whole takeover transition. The findings can complement existing understanding of drivers' behavioral responses to TORs and have important implications on the design of in-vehicle monitoring and alert systems. 



\section{2. METHOD}

\subsection{2.1. Participants}
A total number of 102 university students participated in the study (mean age = 22.9, standard deviation [SD] = 3.8; range = 18-38; 40 females and 62 males). All participants had normal or corrected-to-normal vision and a valid driver license. On average, participants have held their driver license for 4.9 years (SD = 3.2 years). Each participant received a compensation of \$30 for about an hour of participation. 
A 5-point Likert scale was used to measure participants' experience with various driver assistance features (1 indicates “never” and 5 indicates “always”). 
Table \ref{tab:driving experience} showed participants' distribution of annual mileage and weekly mileage, as well as their average experience score with different driver assistance systems.

\begin{table}[H]
\footnotesize
\centering
\caption{Participants' distribution of annual mileage and weekly mileage and average experience score with different driver assistance systems}
\vspace{-2mm}
\captionsetup{justification=centering}
\begin{tabular}{p{3.5cm}p{0.7cm}p{3.5cm}p{0.7cm}p{3.9cm}p{1cm}c} 
\hline
Annual mileage   &  N  & Weekly mileage    &  N   & Driving assistance system       &  Score    \\ \hline
Less than 5,000 miles      & 34      & Less than 50 miles     & 53  & Cruise control & 3.0    \\
5,000 - 10,000 miles       & 33      & 50 - 100 miles    & 27   &  Adaptive cruise control & 1.5  \\
10,000 - 15,000 miles      & 25      & 100 - 150 miles          & 8  & Lane-departure warning & 1.8     \\
15,000 - 20,000 miles      & 2       & 150 - 200 miles          & 6 & lane-keeping assistance & 1.5       \\
20,000 - 25,000 miles      & 5       & 200 - 250 miles          & 8 & Collision warning & 1.9       \\ 
More than 25,000 miles & 3 & More than 250 miles & 2  &
Emergency braking & 1.4\\ 
\hline
\end{tabular}
\label{tab:driving experience}
\end{table}

\subsection{2.2. Apparatus and stimuli}
\begin{figure}[H]
\centering
\includegraphics[height=2in, width=2.8in]{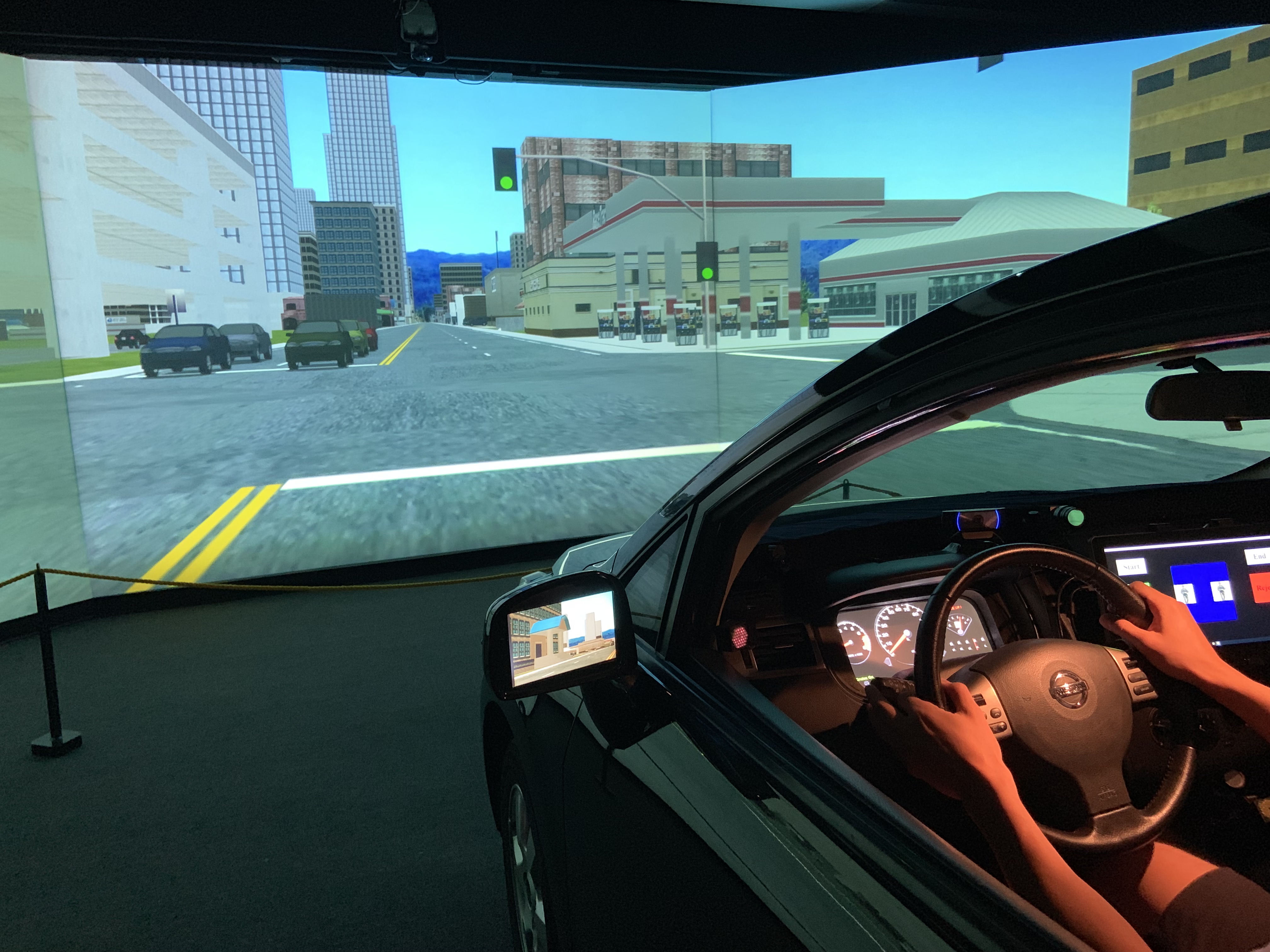}
\caption{The RTI fixed-base driving simulator.} 
\label{fig:drivingsimulator}
\vspace{-5mm}
\end{figure}

The study was conducted in a fixed-base driving simulator from Realtime Technologies Inc. (RTI, Michigan). The virtual world was projected on three front screens (16 feet away), one rear screen (12 feet away), and two side mirror displays (See Figure \ref{fig:drivingsimulator}). 
There was a steering wheel and pedal system embedded in a Nisan Versa car model. The vehicle was programmed to simulate an SAE Level 3 automation, which handled the longitudinal and lateral control, navigation, and responded to traffic events. Participants could press the button on the steering wheel to activate the automated mode and engage in NDRTs. However, the automated mode would be deactivated automatically for drivers to take over control once the automated system failed to respond properly. At that moment, drivers would be alerted by an auditory warning “Takeover”.


\begin{figure}[H]
\centering
\includegraphics[height=1.1in, width=4.1in]{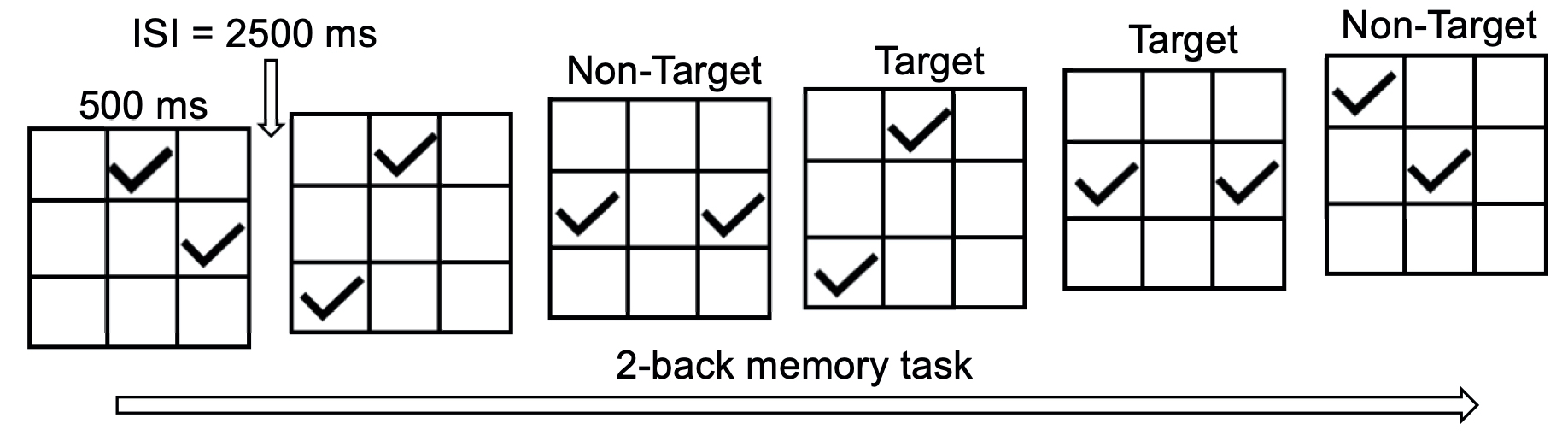}
\caption{N-back memory task.} 
\label{fig:NDRTs}
\vspace{-5mm}
\end{figure}

The NDRT utilized in the study was a visual N-back memory task \citep{jaeggi2008improving}. The stimulus consisted of nine ($3 \times 3$) squares with two human figures randomly appearing in two out of the nine squares. Each stimulus was presented for 500 ms in sequence with a 2500 ms interval (Figure \ref{fig:NDRTs}). Participants were required to press the “Hit” button when the current stimulus was the same as the one presented N steps back in the sequence and press the “Reject” button otherwise. With different N values (i.e. 1 and 2), participants were exposed to conditions with different cognitive load but the same manual and visual load. The task was running on an 11.6-inch touch screen tablet mounted in the center console of the vehicle. 

\begin{figure} [H]
	\centering
	\subfloat[\label{fig:eyetracking}]{\includegraphics[width=0.46\linewidth]{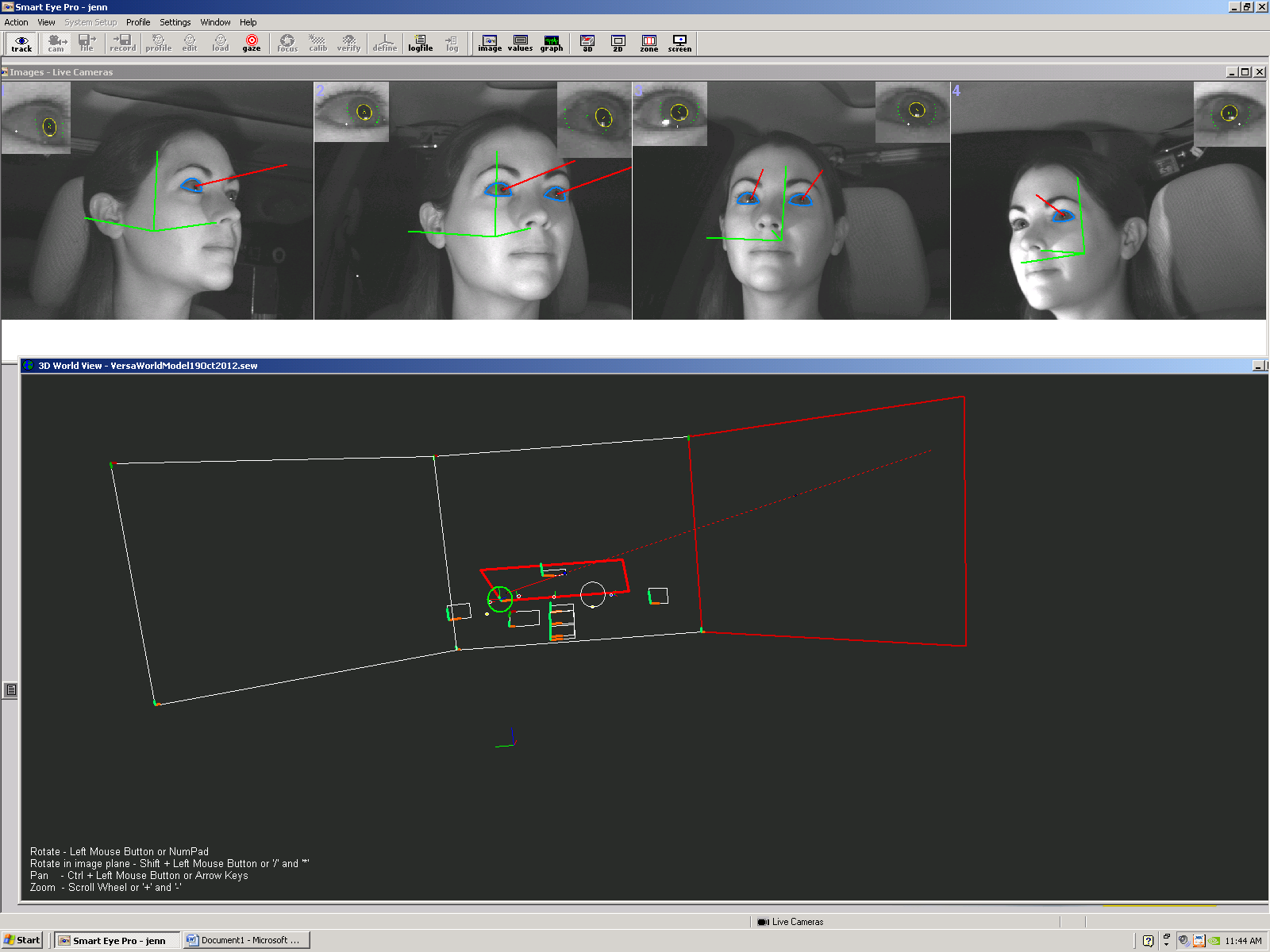}}
	\hspace{0.1cm}
	\subfloat[\label{fig:sensors}]{\includegraphics[width=0.49\linewidth]{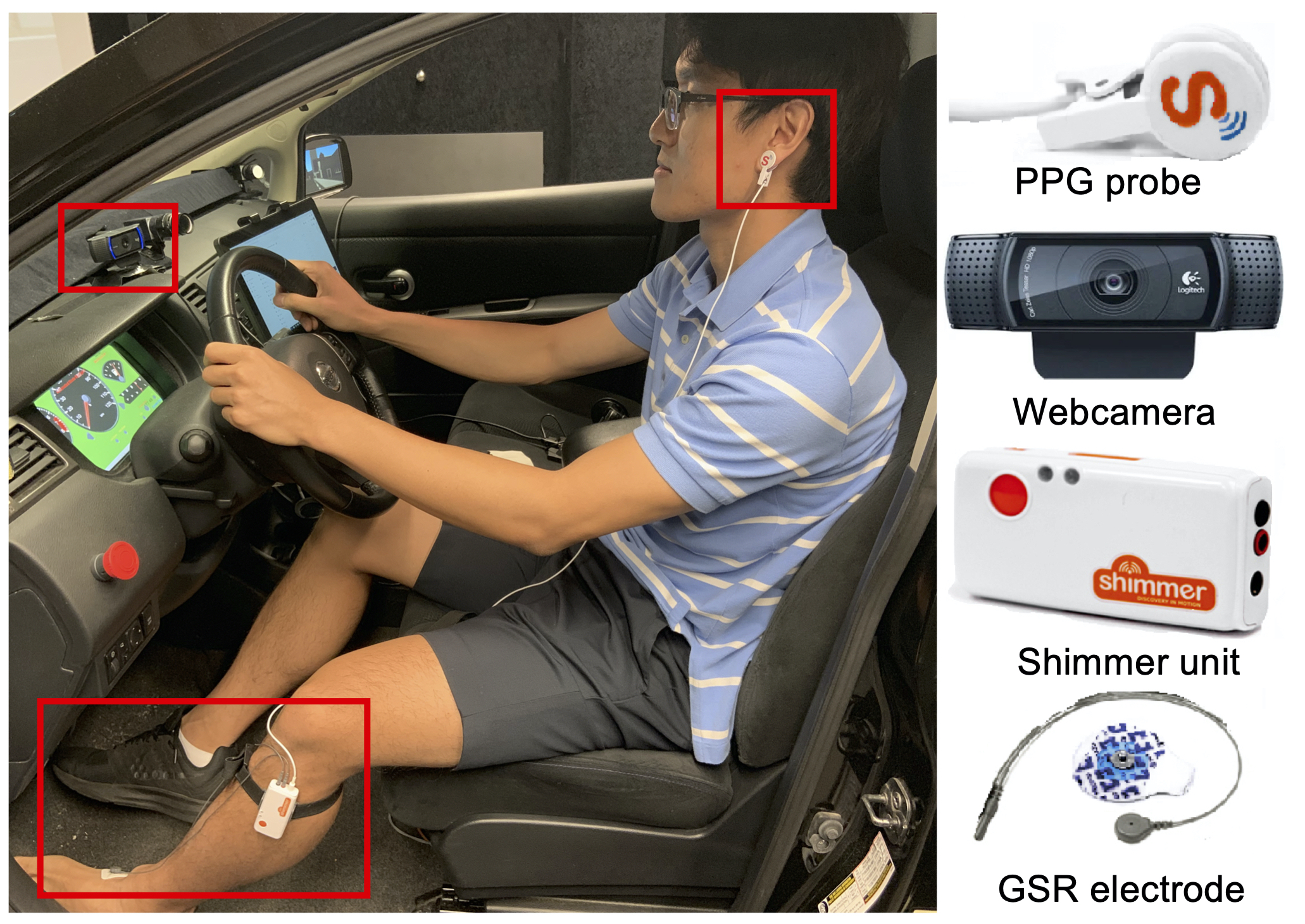}}
	\caption{(a) Smarteye. (b) Shimmer3 GSR+ unit and Logitech web camera.}
	\label{fig:sensors}
\end{figure}

This simulator was equipped with a Smart Eye four-camera eye-tracking system (Smart Eye, Sweden) that provided live head-pose, eye-blink, and gaze data (Figure \ref{fig:eyetracking}). The sampling rate of the eye-tracking system was 120 Hz. The Shimmer3 GSR+ unit (Shimmer, MA, USA) including GSR electrodes and photoplethysmographic (PPG) probe was used to collect GSR and HR data with a sampling rate of 128 Hz. A Logitech web camera with a sampling rate of 30Hz was used to collect drivers' facial expressions (Figure \ref{fig:sensors}). The iMotions software (iMotions, MA, USA) was used for psychophysiological data synchronization and visualization in real time.


\subsection{2.3. Experimental design}
This study employed a within-subjects design with drivers' cognitive load, traffic density, and TOR lead time as independent variables. The cognitive load was manipulated via the difficulty of the NDRTs (low: 1-back memory task; high: 2-back memory task). There were respectively 15 and 0 oncoming vehicles per kilometer  in heavy and light traffic conditions \citep{gold2016taking}. The TOR lead time was 4 or 7 seconds \citep{eriksson2017takeover}. Based on prior literature \citep{koo2016understanding,miller2016behavioral,molnar2018understanding,rezvani2016towards}, eight takeover events were designed in urban and rural drives with typical roadway features: 1) bicyclists ahead; 2) construction zone on the left; 3) construction zone ahead; 4) sensor error on the right curve; 5) swerving vehicle ahead; 6) no lane markings on the curve; 7) sensor error on the left curve; 8) police vehicle on shoulder. 
The order of cognitive load, traffic density, and TOR lead time was counterbalanced via an $8 \times 8$ balanced Latin Square across participants. Considering standard programming practices for the simulator, the order of scenario presentations was counterbalanced by having half of the participants drive from Event 1 to 8, and the other half from Event 8 to 1 \citep{bingham2016peer}. There were no other vehicles in the driver’s direction so the participants could avoid the objects in their lane by changing to the adjacent lane. The AV was always in the right lane prior to the TOR. 

\subsection{2.4. Dependent measures}
We collected drivers' psychophysiological measures, vehicle-related measures, and subjective ratings of takeover performance in the study. Vehicle-related results were not reported in this paper. The psychophysiological measures included drivers' gaze behaviors, HR activities, GSRs, and facial expressions. All the dependent variables were summarized in Table \ref{tab:dependent variables}. 

PPG peaks were detected using an adaptive threshold method for heart rate extraction \citep{shin2009adaptive}. Heart rate variability was calculated as the standard deviation of RR intervals (i.e., the time elapsed between two successive R-waves on the electrocardiogram) \citep{castaldo2017heart}. 
In addition to directly measuring drivers' average heart rate in takeover stages relative to the NDRT stage, we also categorized such heart rate differences into three patterns because it can reflect drivers' attentional styles during transitions as introduced before.
Heart rate acceleration/deceleration was defined as at least 2 heart beats per minute (bpm) increase/decrease from the NDRT stage to the takeover stage. No changes in heart rate indicated less than 2 bpm changes between two stages \citep{pohlmeyer2008association,reimer2011impact}.

The raw GSR signals were decomposed into phasic and tonic components using the continuous decomposition analysis (CDA) via Ledalab in Matlab \citep{benedek2010continuous}.
Then maximum and mean phasic components were calculated for further analysis as they were responsible for relatively rapid changes in response to specific events in the takeover transitions \citep{wintersberger2018let}. For gaze behaviors, we calculated drivers' eyes-on-road time percentage, blink number, and horizontal gaze dispersion. Horizontal gaze dispersion was defined as the standard deviation of gaze heading. Drivers' emotional valence and engagement were extracted from their facial expressions using iMotions Affectiva module to reflect how positive/negative and expressive their emotions were \citep{stockli2018facial,kulke2020comparison}.

\begin{figure}[H]
\centering
\includegraphics[height=1.8in, width=5in]{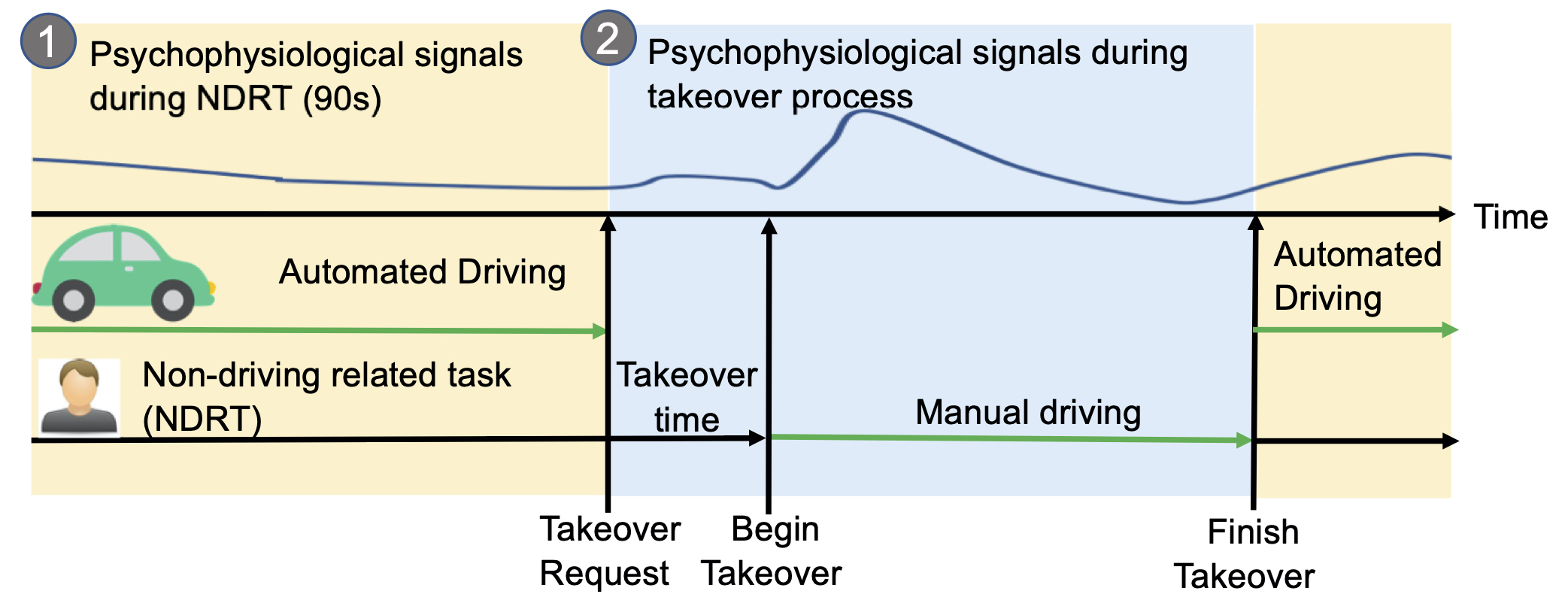}
\caption{Two time windows (see corresponding results in Subsection 3.1 and Subsection 3.2) to calculate measures from psychophysiological signals.} 
\label{fig:physiological_twowindow}
\vspace{-5mm}
\end{figure}

We calculated the above-mentioned statistical measures using two time windows: the NDRT process and the takeover process (see Figure \ref{fig:physiological_twowindow}). The NDRT process was approximately 90-second long and was started when the drivers were asked to initiate the NDRT and ended when the auditory “Takeover” alert was issued. The takeover stage started with “Takeover” alerts and ended when drivers negotiated takeover events and re-engaged the vehicle. In order to show the continuous takeover transition process, we also depicted the psychophysiological measures after TORs second by second when their main effects were significant.


\begin{table}[H]
\small
\centering
\caption{Dependent Variables.}
\label{tab:dependent variables}
\begin{tabular}{p{4cm}p{1.7cm}p{3.3cm}p{4.7cm}c}
\hline
\multicolumn{1}{l}{Dependent measures} & \multicolumn{1}{l}{Unit} & \multicolumn{1}{l}{Category} & \multicolumn{1}{l}{Explanation} \\ \hline
Heart rate variability &  millisecond & Heart rate &  Standard deviation of inter-beat-interval \\
 Difference in average heart rate &  beat per minute & Heart rate & Difference in average heart rate between NDRT and takeover stage \\
 Mean phasic GSR  & micro Siemens & GSR &  Average GSR phasic activation \\
Maximum phasic GSR  & micro Siemens  & GSR &  Maximum GSR phasic activation\\
Eyes-on-road time & percentage & Gaze behaviors & The time percentage while eyes are on the road\\
Blink number &  &  Gaze behaviors & The number of blinks \\
Horizontal gaze dispersion &  radian & Gaze behaviors &  The standard deviation of gaze heading \\
Emotional valence &  -100 to 100  & Facial expressions & Signs indicate positive or negative emotions \\
Emotional engagement &  0 to 100  & Facial expressions & Increasing values signify increased emotional engagement \\
Takeover performance &  0 to 100  & Subjective rating & Larger values indicate better self-reported takeover performance \\
\hline
\end{tabular}%
\label{tab:measures}
\end{table}

\subsection{2.5. Experimental Procedure}
The participants were first briefed about the study. After participants signed an informed consent form and completed an online demographics questionnaire, they were asked to track six targets on the front screen for eye-tracking calibration. Next, two GSR electrodes were attached to their left foot and the PPG probe to the left ear lobe. Participants were informed that there was no need to actively monitor the driving environments or take over control of the vehicle as long as no TOR was issued since the vehicle was able to handle the situations itself.

\begin{figure}[H]
\centering
\includegraphics[height=1.5in, width=4.6in]{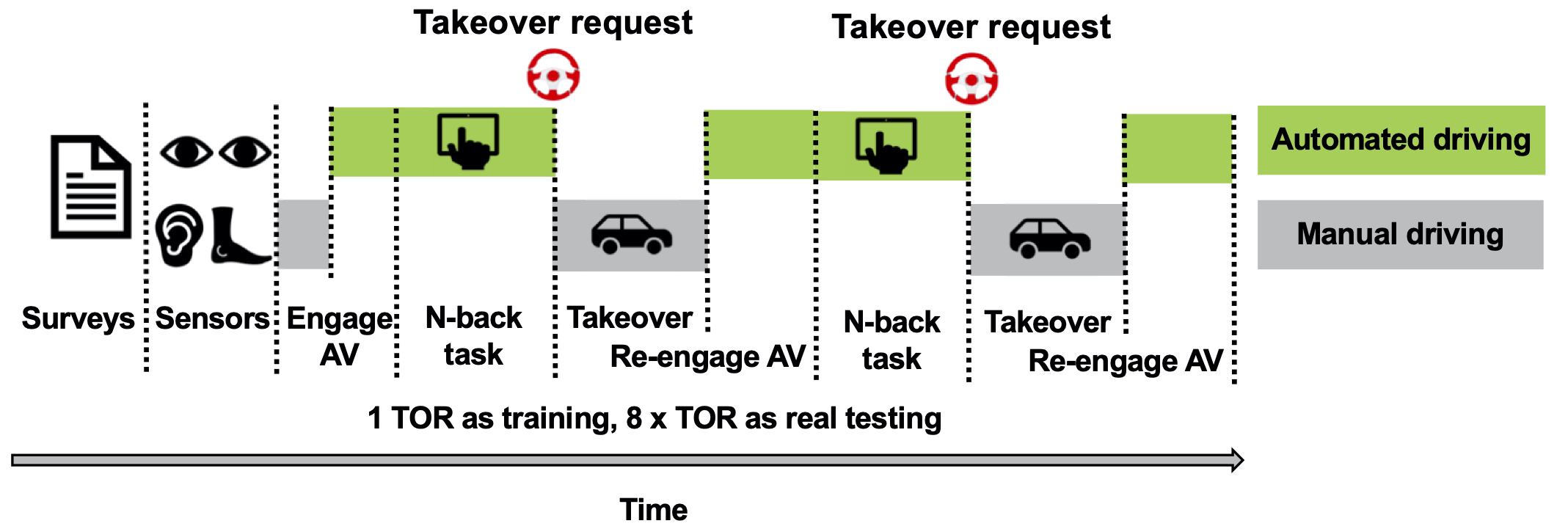}
\caption{Experiment procedure.} 
\label{fig:procedure}
\vspace{-5mm}
\end{figure}

Participants had a 2-minutes practice for the N-back memory task, followed by a 5-minutes practice drive to get familiar with the simulator environment. Participants were informed that they would get additional 20 dollars if their NDRT performance in the real experiment was ranked among top 10. Next, each participant drove two experimental drives (10-20 minutes each), each containing four takeover events. At the beginning of the drive, participants were asked to activate the AV mode and then start the N-back task when the audio command “Please start the NDRT” was issued. After about 90-second NDRT, a TOR was issued unexpectedly, and participants were required to terminate the NDRT manually and take over the control immediately. When participants thought they had negotiated the takeover event, they were free to activate AV mode and were not encouraged to keep driving all the time. The operation of NDRT, takeover, and AV mode activation were repeated for each takeover event (Figure \ref{fig:procedure}). There was a break stage between each repetition and the experimenter would make sure that participants were in the AV mode when the next NDRT command was issued.
After each takeover event, participants reported their takeover performance for each takeover event using a visual analogue scale, with 0 indicating not good at all and 100 indicating very good. The survey on takeover performance was administered on a touch screen after each takeover event with AV mode activated. 

\subsection{2.6. Data analysis}
Each participant experienced 8 scenarios, so 102 participants yielded a total of 816 ($8 \times 102$) scenarios. Due to some participants' motion sickness and malfunctions of driving simulator and psychophysiological sensors (e.g., calibration failure of steering wheel and eye-tracking system, system freezing), 683 scenarios were available for further analysis. 

Two types of linear mixed models were conducted using SPSS version 24 to examine effects on continuous dependent variables (Table \ref{tab:dependent variables}). The first one used cognitive load, TOR lead time, traffic density, and their interactions as fixed effects and the second one used time window (NDRT process vs. takeover process) as fixed effect. 
Subjects were treated as random effects to resolve non-independence in all the models. 
Levene’s tests were conducted to examine the assumption of homogeneity of variance. All the dependent variables showed equal variance across the cognitive load, traffic density, and TOR lead time levels. Although the Shapiro–Wilk tests showed that the assumption of normality was violated for some dependent variables (e.g., horizontal gaze dispersion), we argued that linear mixed models can still be conducted because they are robust against violations of the assumptions of normality \citep{gelman2006data}. 
Meanwhile, if the main effects of independent variables on psychophysiological measures during the takeover process were significant, we used pairwise  \textit{t}-tests to compare psychophysiological measures after TORs second by second to provide time-series insights. 
Since heart rate change pattern was a categorical variable, we used the chi-squared test to examine its dependence with independent variables, which could represent drivers' attentional styles in different conditions \citep{pohlmeyer2008association,reimer2011impact}. To increase the interpretation of psychophysiological results, Pearson correlation coefficients were examined to explore the relationships between emotions, takeover performance, and other physiological data. The significance level alpha was set at .05. We calculated partial eta squared ($\eta_{\text{p}}^{2}$), Cohen's \textit{d}, and Phi ($\varphi$) as effect sizes for the linear mixed models, \textit{t}-tests, and chi-squared test, respectively \citep{kim2017statistical,lakens2013calculating,cohen1965some}. 

\section{3. RESULTS}
The result section has three parts. Drivers' psychophysiological responses including heart rate variability and gaze behaviors during NDRTs were presented in Subsection 3.1. Subsection 3.2 showed drivers' psychophysiological responses to TORs including gaze behaviors, galvanic skin responses, and heart rate.
Subsection 3.3 demonstrated the correlations between drivers' emotions, takeover performance, and physiological data.

 \subsection{3.1. Psychophysiological responses during NDRTs}

 \subsubsection{Heart rate variability}
\begin{figure}[H]
\centering
\includegraphics[height=2.1in, width=3in]{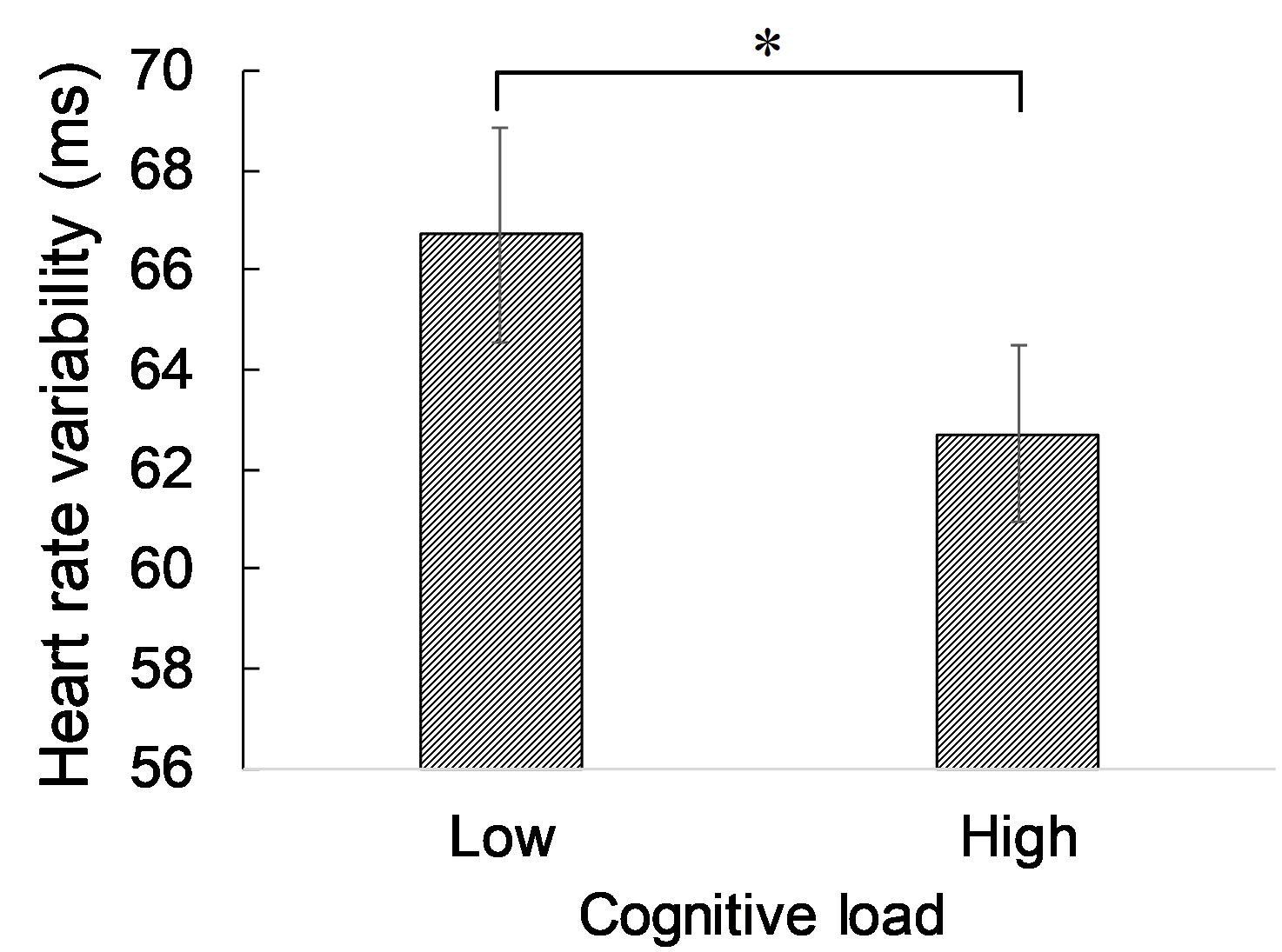}
\caption{Heart rate variability during NDRTs by cognitive load. We use the following indications for all the figures and tables applicable:  ***Difference is significant at the 0.001 level; **Difference is significant at the 0.01 level; *Difference is significant at the 0.05 level. Error bars indicate 1 standard error (SE).} 
\label{fig:HeartRateVariability}
\vspace{-5mm}
\end{figure}

During NDRT, there was a significant main effect of cognitive load on heart rate variability ($F(1, 586)=5.17, p=.023, \eta_{\text{p}}^{2} = .01$). Drivers had lower heart rate variability when they were in the condition of high cognitive load (Figue \ref{fig:HeartRateVariability}). All other main effects and interaction effects on heart rate variability were not significant, so they were not included in Figure \ref{fig:HeartRateVariability}. 

 \subsubsection{Gaze behaviors}
\begin{figure} [H]
	\centering
	\subfloat[\label{fig:gazedispersion}]{\includegraphics[height=0.35\linewidth]{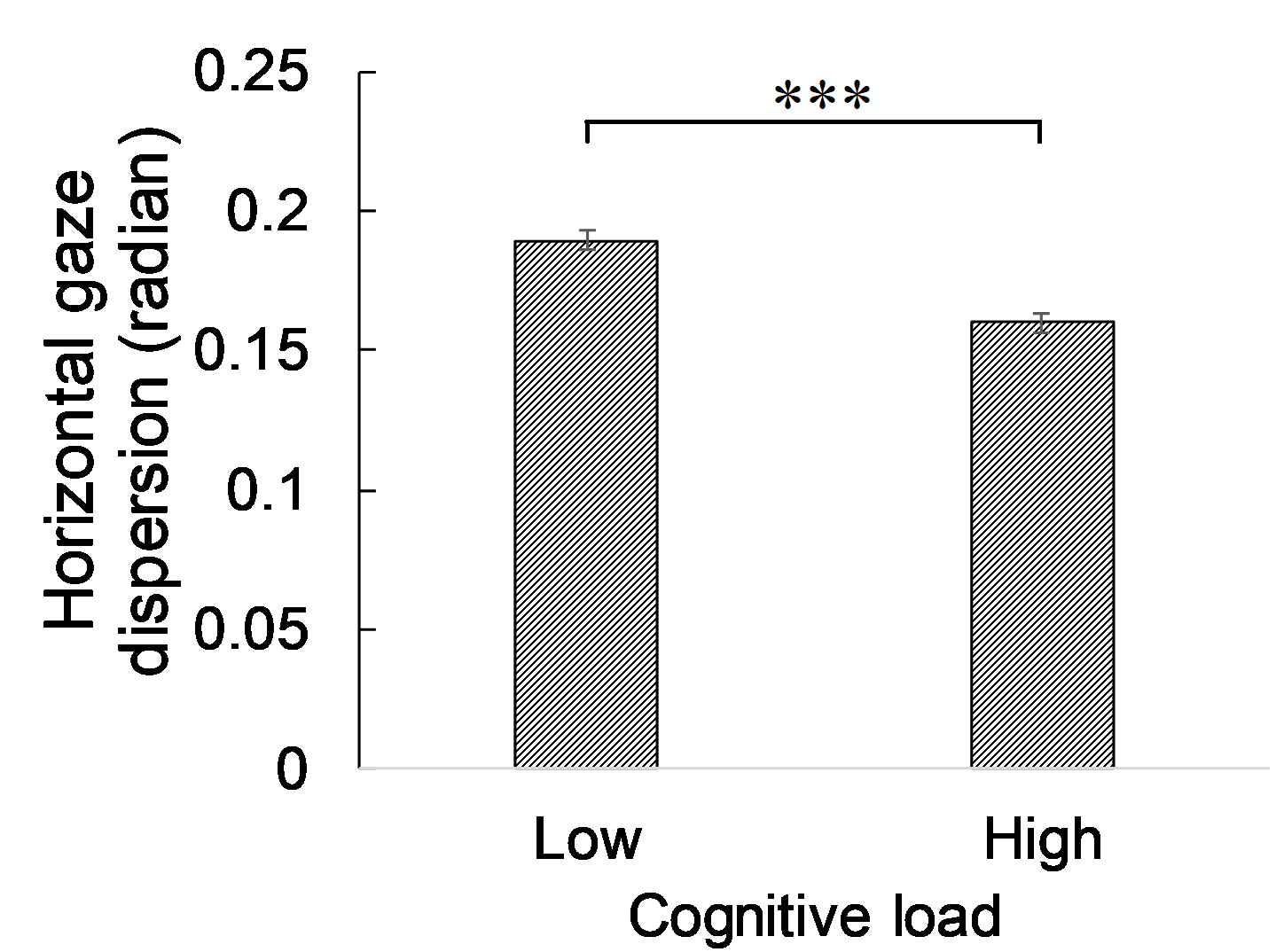}}
	\hspace{0.1cm}
	\subfloat[\label{fig:eyes-on-road}]{\includegraphics[height=0.35\linewidth]{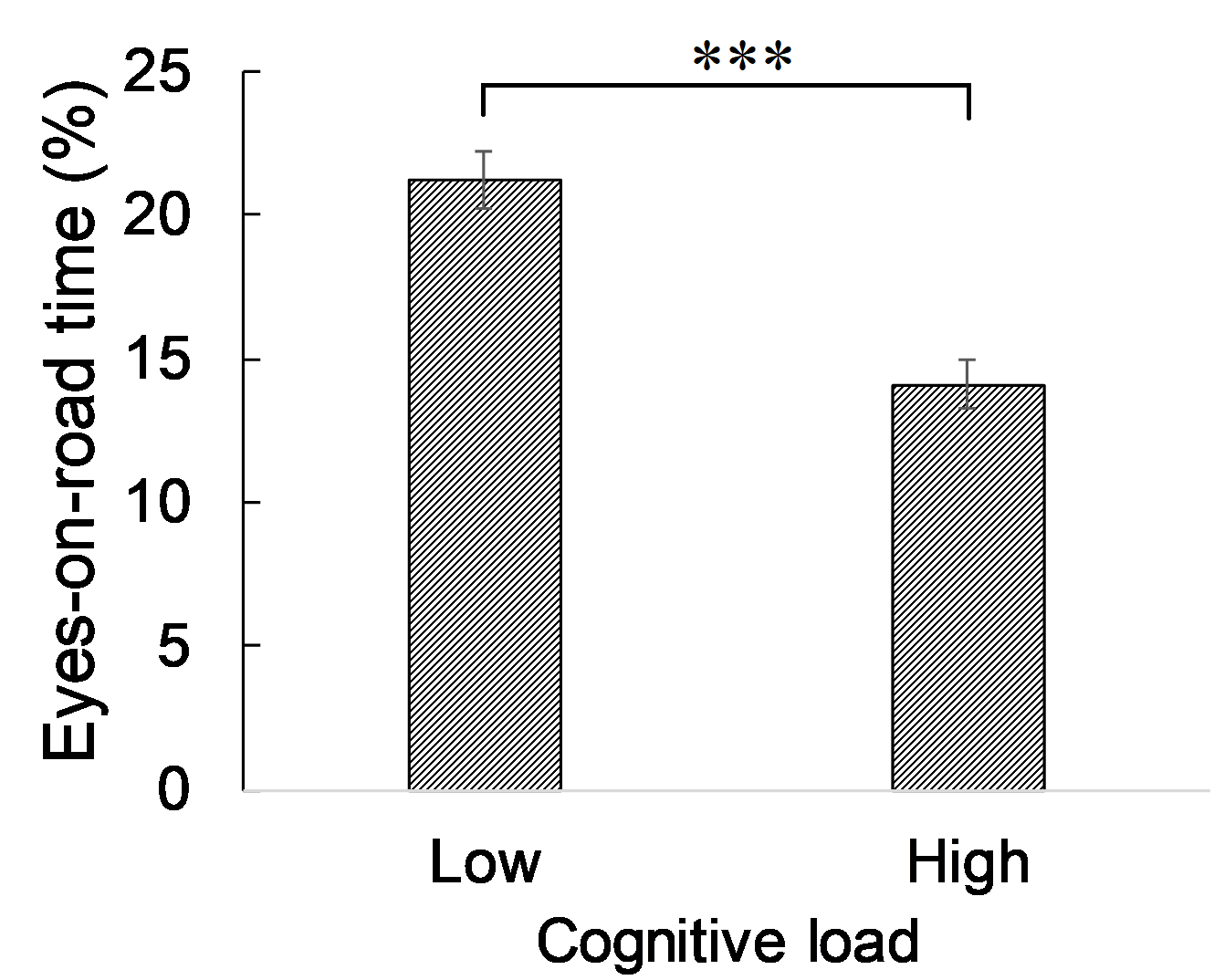}}
	\caption{(a) Horizontal gaze dispersion; (b) Eyes-on-road time percentage during NDRT process by cognitive load. TORs were issued at Time 0.}
	\label{fig:gaze}
\end{figure}

As shown in Figure \ref{fig:gaze}, drivers had lower horizontal gaze dispersion ($F(1, 586)=108.75, p<.001, \eta_{\text{p}}^{2} = .16$) and shorter eyes-on-road time ($F(1, 586)=108.35, p<.001, \eta_{\text{p}}^{2} = .16$) when they were in high cognitive load. However, their blink number did not differ significantly between two cognitive load task conditions. The main effects of traffic density and TOR lead time and their interaction effects were not significant and were not included in the Figure \ref{fig:gaze}. 

 \subsection{3.2. Psychophysiological responses during takeover transitions}
\subsubsection{Gaze behaviors}
\begin{figure}[H]
\centering
\includegraphics[height=2.1in, width=3in]{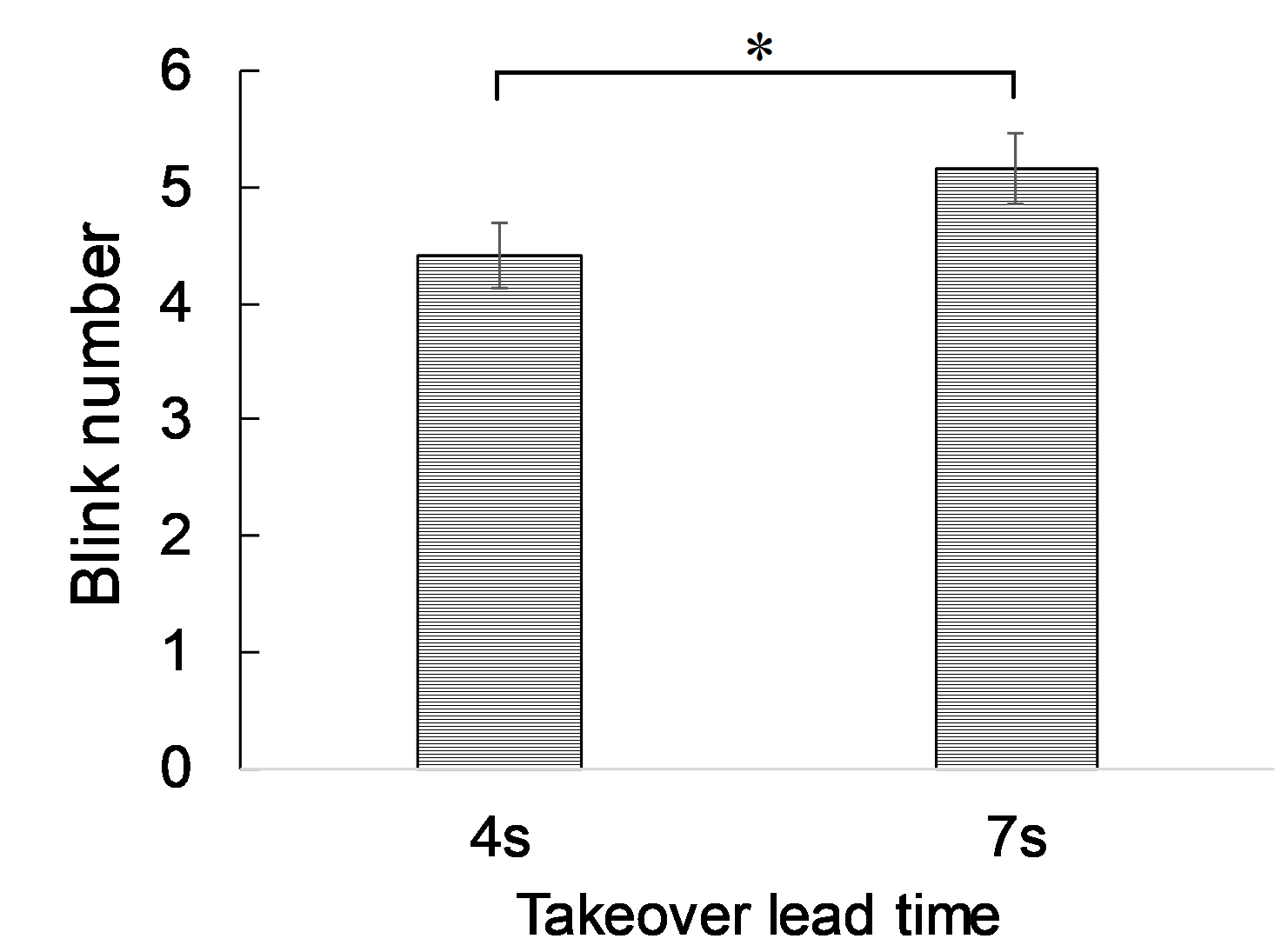}
\caption{Blink number after TORs by TOR lead time.} 
\label{fig:blinknum}
\vspace{-5mm}
\end{figure}

\begin{figure}[H]
\centering
\includegraphics[height=3in, width=4.2in]{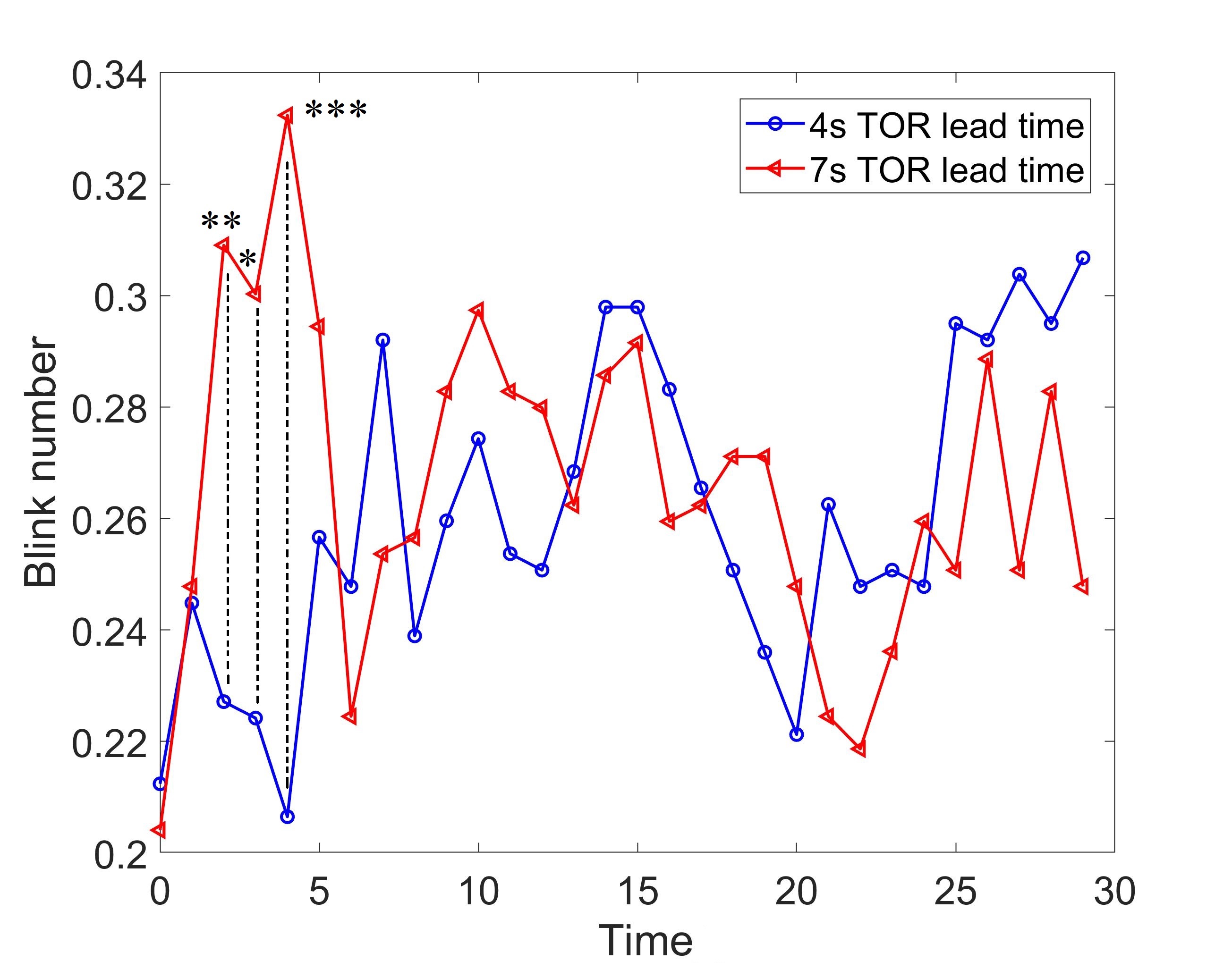}
\caption{Blink number through the drives. TORs were issued at Time 0. } 
\label{fig:BlinkTimeseries}
\vspace{-3mm}
\end{figure}

Only the main effect of TOR lead time on blink number was significant ($F(1, 588)=6.11, p=.014, \eta_{\text{p}}^{2} = .01$). We found that 4s TOR lead time led to fewer blink numbers than 7s TOR lead time in general during takeover process (Figure \ref{fig:blinknum}). If we analyzed the blink number second by second, as shown in Figure \ref{fig:BlinkTimeseries}, we found that 4s TOR lead time significantly suppressed blinks at 2s, 3s, and 4s after TORs (2s: $t(90)=2.96, p=.004, Cohen's$ $d= .31$; 3s: $t(90)=1.78, p=.05, Cohen's$ $d= .19$; 4s: $t(90)=4.51, p<.001, Cohen's$ $d= .48$). 
Yet, no significant effects were found on the horizontal gaze dispersion.

\subsubsection{Galvanic skin responses}
\begin{figure}[H]
\centering
\includegraphics[height=2.9in, width=4.5in]{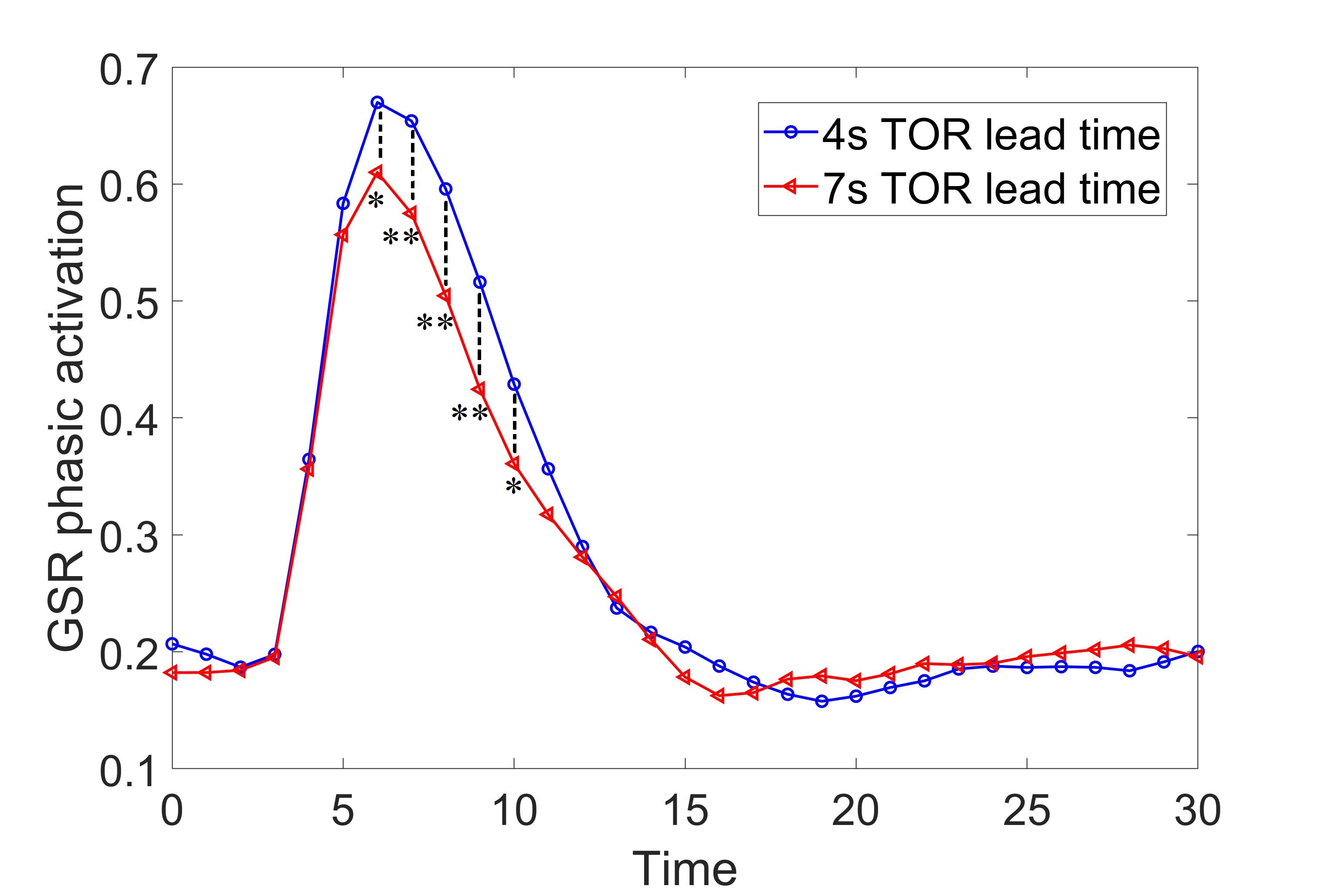}
\caption{Mean GSR phasic through the drives. TORs were issued at Time 0. } 
\label{fig:GSRTimeseries}
\vspace{-3mm}
\end{figure}

Compared to the NDRT stage, drivers' mean phasic GSR was significantly higher in the takeover action stage ($F(1, 1275)=44.43, p< .001, \eta_{\text{p}}^{2} = .03$). 
As shown in Figure \ref{fig:GSRTimeseries}, drivers' GSR phasic activation increased after a TOR and reached a peak 5s after the alert. The main effects of TOR lead time on maximum and mean GSR phasic activation were significant ($F(1, 587)=8.80, p=.003, \eta_{\text{p}}^{2} = .01$; $F(1, 591)=4.92, p=.027, \eta_{\text{p}}^{2} = .01$).  Generally, 4s TOR lead time induced larger maximum and mean GSR phasic activation than 7s TOR led time during the whole takeover time window. Furthermore, we found that GSR phasic activation differences caused by TOR lead time appeared 5s after the TOR, lasted for 5s and disappeared 10s after the TOR (5s: $t(90)=2.33, p=.022, 
Cohen's$ $d= .25$; 6s: $t(90)=2.87, p=.005, Cohen's$ $d= .30$; 7s: $t(90)=3.20, p=.002, Cohen's$ $d= .34$; 8s: $t(90)=3.14, p=.002, Cohen's$ $d= .33$; 9s: $t(90)=2.43, p=.017, Cohen's$ $d= .26$). No other significant effects were found on the mean or maximum GSR phasic activation.



 \subsubsection{Heart rate}
 
\begin{figure}[H]
\centering
\includegraphics[height=1.8in, width=4.8in]{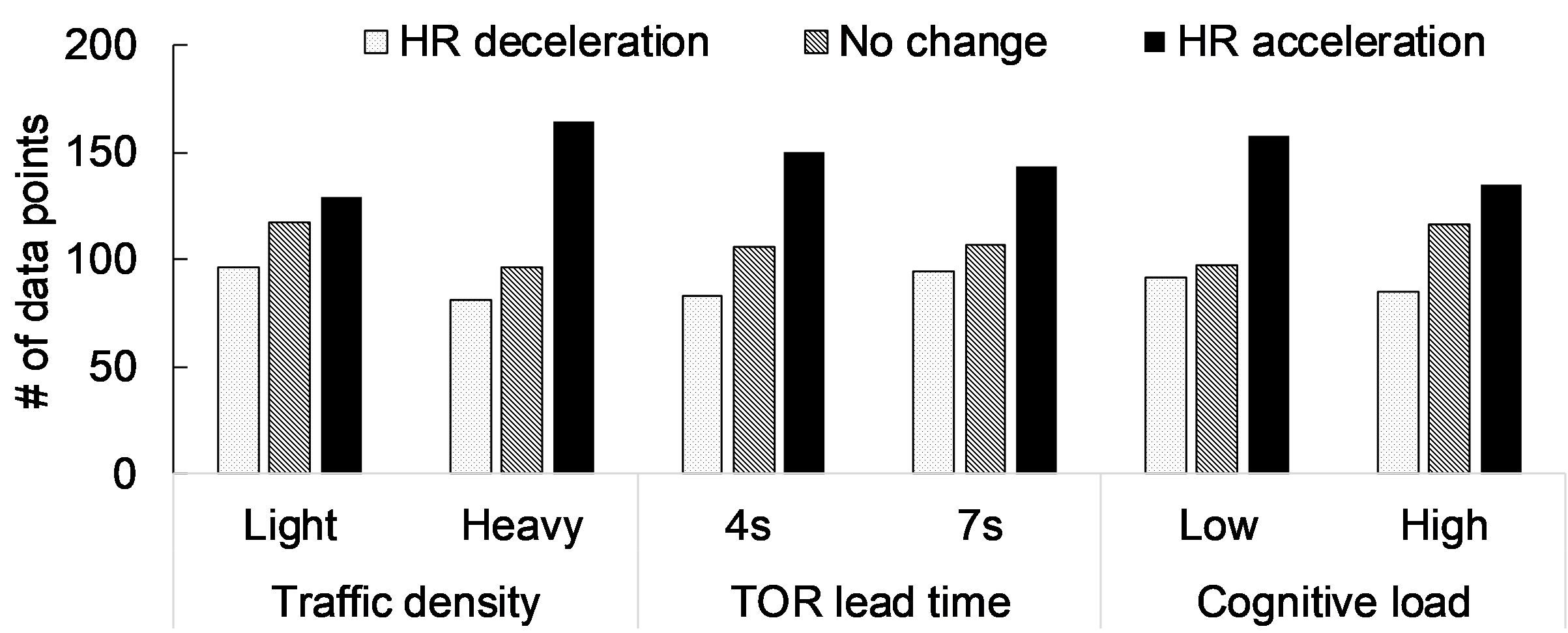}
\caption{The number of takeover scenarios by independent variables and HR response pattern.} 
\label{fig:HRChangesNum}
\vspace{-5mm}
\end{figure}

\begin{figure}[H]
\centering
\includegraphics[height=3in, width=4.1in]{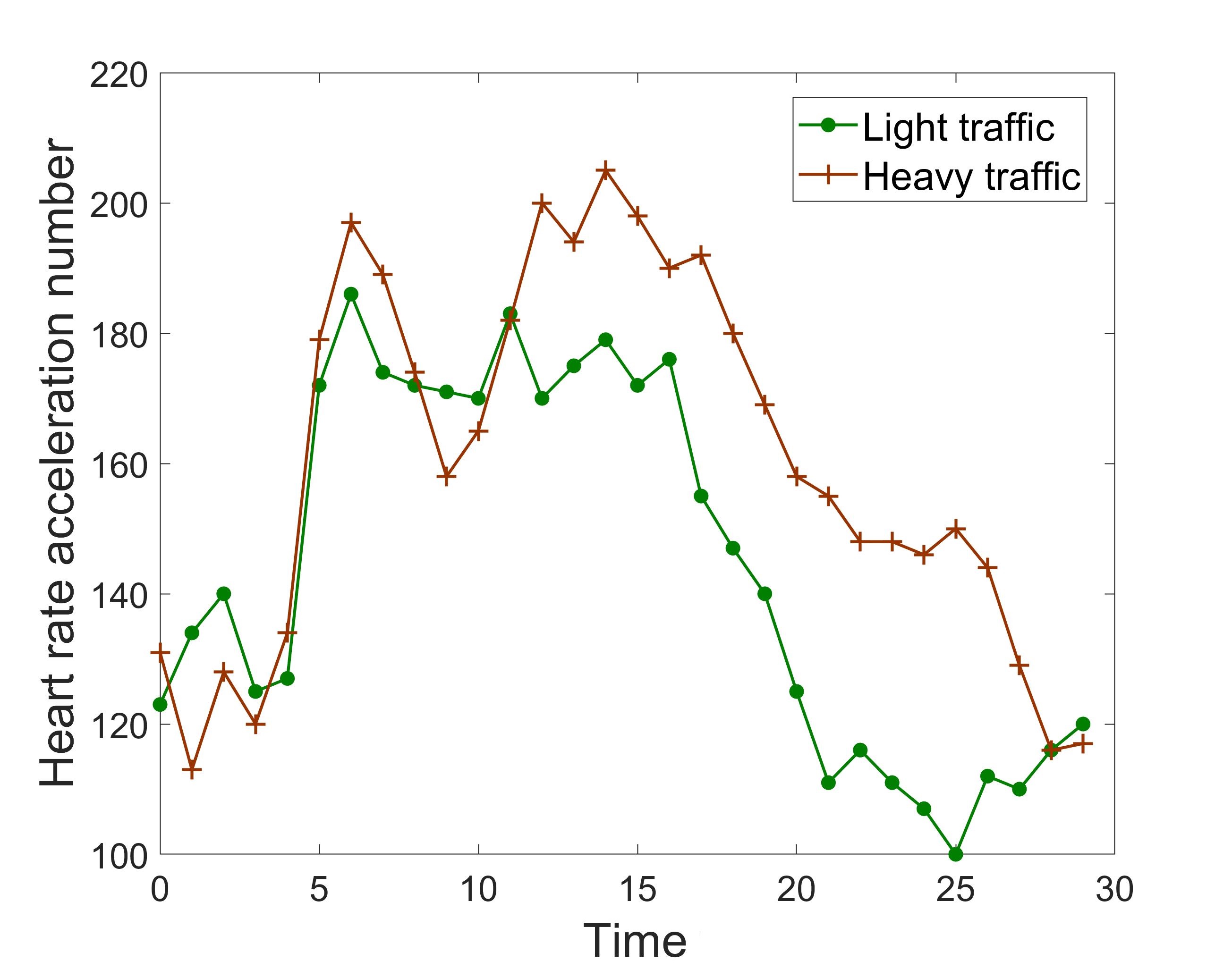}
\caption{The number of heart rate acceleration patterns after TORs. TORs were issued at Time 0.} 
\label{fig:HRPTimeseries}
\vspace{-3mm}
\end{figure}

The main effects of cognitive load, traffic density, TOR lead time, and their interaction effects on exact values of heart rate changes (heart rate in the takeover stage minus NDRT stage) were not significant. As introduced in Subsection 2.4, heart rate differences were then categorized into three patterns.
Figure \ref{fig:HRChangesNum} shows the number of three heart rate response patterns under different traffic density, TOR lead time and  cognitive load conditions. Primarily, heart rate acceleration happened the most frequently when drivers switched from NDRTs to takeovers, followed by no changes, and heart rate deceleration. There was a significant main effect of traffic density on heart rate response patterns (${\chi}^2_{2}=7.54, p=.023, \varphi = .11$). In comparisons to light traffic density, significantly more heart rate acceleration patterns were found in the heavy traffic density condition (Table \ref{tab:HRchangesTraffic}). As shown in Figure \ref{fig:HRPTimeseries}, such differences appeared 12th second after TORs and lasted until about 27th second. 
Yet, the main effects of TOR lead time and cognitive load on heart rate response patterns were not significant. 


 \begin{table}[H]
 \small
\centering
\caption{Mean heart rate (and standard error) by traffic density group and HR response pattern.}
\label{tab:HRchangesTraffic}
\begin{tabular}{p{0.8cm}p{1.7cm}p{1.7cm}p{1.7cm}p{1.7cm}p{1.5cm}p{1.7cm}c} 
\hline
\multicolumn{1}{c}{\multirow{2}{*}{Stage}} &
  \multicolumn{3}{c}{Light traffic density} &
  \multicolumn{3}{c}{Heavy traffic density} \\ \cline{2-7} 

 
 & HR deceleration (n = 96) &
 No changes (n = 117)&
 HR acceleration (n = 129) &
  HR deceleration (n = 81) &
 No changes (n = 96)&
 HR acceleration (n = 164) \\ \hline
 NDRT &
  $92.1\pm{3.1}$ &
  $80.3 \pm{1.3}$&
  $81.2 \pm{1.8}$&
    $90.0 \pm{2.4}$&
$80.8 \pm{2.0}$&
  $81.6 \pm{1.6}$\\ 
\multicolumn{1}{l}{Takeover} &
  \multicolumn{1}{l}{$85.2\pm{2.6}$} &
  \multicolumn{1}{l}{$80.5\pm{1.3}$} &
  \multicolumn{1}{l}{$91.1\pm{2.4}$} &
    \multicolumn{1}{l}{$83.0\pm{2.0}$} &
      \multicolumn{1}{l}{$80.9\pm{1.9}$} &
  \multicolumn{1}{l}{$88.6\pm{1.8}$} \\ \hline
\end{tabular}%
\end{table}

 \subsection{3.3. Correlations Matrix} 
The correlation matrix, shown in Table \ref{tab:correlation}, indicates the relationships between drivers' physiological data, subjective ratings of performance, and emotions in valence and engagement dimensions after TORs. We found that maximum and mean GSR phasic activation were negatively correlated with drivers' emotional valence, whilst blink number was positively correlated with drivers' emotional valence.
In other words, the more negative emotions drivers had, the larger maximum and mean GSR phasic activation and less blink number they had after TORs. Meanwhile, drivers' engagement was significantly positively correlated with HR differences between takeover and NDRT stage, while subjective ratings of takeover performance were significantly negatively correlated with horizontal gaze dispersion.

\begin{table}[H]
 \small
\centering
\caption{Correlations Matrix between drivers' physiological data, emotions, and subjective takeover performance.}
\label{tab:correlation}
\begin{tabular}{p{2.2cm}p{2.8cm}p{1cm}p{2cm}p{2cm}p{2cm}c}
\hline
&  Horizontal gaze dispersion &  Blink num &  HR differences &  Max GSR phasic  &  Mean GSR phasic  \\\hline
Valence     & -.042     & .123**    & -.002   & -.158** & -.107** \\
Engagement     & -.017  & .051    & .09*    & -.042 & -.069 \\
Performance     & -.092*  & 0    & -.051    & -.042 & -.055 \\
\hline
\end{tabular}%
\end{table}

\section{4. DISCUSSION}

This exploratory study examined the effects of NDRTs, traffic density, and TOR lead time on drivers' psychophysiological responses to TORs in simulated SAE Level 3 automated driving. The systematic analysis of psychophysiological measures gave us an overview of drivers' cognitive and emotional states, attention, and situational awareness throughout the whole takeover process both at the overall level and at the continuous level.

\subsection{4.1. Psychophysiological measures during NDRTs}
During the NDRT stage with automated driving mode on, drivers were assigned N-bask tasks on the tablet. Our results showed that drivers had lower heart rate variability when they were in 2-back memory task than 1-back memory task. Heart rate variability is a sensitive indicator of cognitive load \citep{mehler2012sensitivity,lei2011influence}. Our findings aligned with previous research \citep{bashiri2014heart,mehler2011agelab}, and implied drivers' high cognitive load in 2-back memory task. 

Meanwhile, we found that drivers had narrower horizontal gaze dispersion and spent less time monitoring the road when they were in 2-back memory task. This can be explained from two aspects. First, 2-back memory task required drivers to memorize more chucks and required more cognitive resources. Consistent with previous studies \citep{wang2014sensitivity,gold2016taking}, narrower horizontal gaze dispersion indicated drivers' increased cognitive load in 2-back memory task.
Second, while more attentional resources were occupied by the 2-back memory task, drivers had fewer opportunities to monitor the driving environment. Their narrower horizontal gaze dispersion and less time of eyes on road suggested reduced situational awareness of the driving environment \citep{molnar2017age}.

\subsection{4.2. Psychophysiological measures during takeover transitions}

Upon the TOR, drivers were required to terminate NDRTs, check the driving environment, and negotiate takeover scenarios appropriately. During this process, we found that drivers had fewer blink numbers when TOR lead time was 4s. The number of blinks decreases when there is more information to be processed in a short period of time \citep{veltman1996physiological}. Thus, blink inhibition in 4s TOR lead time indicated that drivers paid greater attention to scenarios and utilized more efforts to support decision making and respond to urgent events. Meanwhile, we found that blink number was positively correlated with drivers' emotional valence detected by facial expressions. This suggested that the more blink suppression drivers had, the more negative emotions (e.g., stress) drivers had in the face of TORs
\citep{haak2009detecting}. However, we did not find significant differences of blink number in two different cognitive load conditions. This was probably because blink number was more sensitive to temporal demands \citep{veltman1996physiological} than to cognitive demands. Meanwhile, we found a significantly negative correlation between drivers' subjective ratings of takeover performance and horizontal gaze dispersion. It is likely that drivers required wider horizontal gaze dispersion to process the driving information and negotiate takeover events in a worse takeover performance situation.

Regarding GSRs, drivers' phasic components increased significantly in response to TORs, which implies high emotional arousal to unexpected events \citep{boucsein2012electrodermal}.
In general, compared to 7s TOR lead time, drivers had larger maximum and mean GSR phasic activation in the 4s TOR lead time condition, indicating higher arousal when situations were more critical. However, a high arousal level could both be associated with positive and negative emotions. Therefore, we further looked into its correlation with drivers' emotional valence.
We found that maximum and mean GSR phasic activation were negatively correlated with drivers emotional valence. In other words, the higher arousal the drivers had in response to TORs, the more negative the drivers' emotions were. Following the previous studies
\citep{wandtner2018effects,morris2017electrodermal,healey2005detecting}, we interpreted that drivers experienced greater stress in the 4s TOR lead time condition as indicated by the GSR phasic component and emotional valence. 


As described in the results section, there were different patterns of drivers' average heart rate differences from NDRTs to takeover stage. In general, heart rate acceleration happened the most frequently, which was associated with stimulus ignorance and environmental rejection \citep{sokolov1963higher,lacey1967somatic,lacey1970some}. Such an attentional pattern matched the takeover mechanism as drivers were required to terminate or ignore their NDRTs for takeover actions at the moment of TOR. More interestingly, compared to light traffic density, drivers showed more heart rate acceleration patterns in heavy traffic density. This meant that drivers selectively rejected and blocked out of the overwhelmed traffic information in attention-demanding situations. Even though we did not find any performance-level differences induced by traffic density \citep{du2020autoui}, heart rate measures explained drivers' attentional styles and revealed potential safety concerns with heavy traffic density during takeover transitions. 
Also, there was a significant positive correlation between drivers' engagement and heart rate changes from NDRT to takeover stages. The more heart rate acceleration drivers had, the more engaged they were in the takeover transitions, indicating that drivers were engaged in takeover actions while ignoring unnecessary traffic information in complex situations indicated by heart rate acceleration patterns.



\subsection{4.3. Time-series psychophysiological measures}
The second-by-second analysis of psychophysiological measures allow us to understand drivers' responses to TORs in a continuous way. 
Using time-series data, we found that drivers' blink suppression happened 2s after TORs and lasted for 3 seconds. The onset of the significant differences at the 2nd second tended to be consistent with drivers' reaction time (average reaction time = 2.3 s in this study)  \citep{eriksson2017takeover,mcdonald2019toward}. Once drivers started to take over control of the vehicle, their blinks were suppressed to extract the most important visual information and remove distracting information in the driving environment \citep{bidder1997comparison}. Yet, compatible with the characteristics of gaze behaviors in previous studies \citep{alrefaie2019heart,kramer2013processing}, such gaze reactions to TORs were rapid and could recover immediately when the complex driving information was processed.

With regard to GSR phasic activation, we found that drivers' phasic differences triggered by different lead times became significant 5s after the TOR, but lasted only for another 5 seconds and then became monotonous. This was likely because drivers perceived the event urgency differently at the time of TOR, but got used to it after they gradually negotiated takeover scenarios. This phenomenon was also consistent with the latency of GSRs responding to unexpected events, the rise time to the peak from the baseline, and the fall time returned to the baseline from the peak after unexpected events were resolved \citep{boucsein2012electrodermal}. 

However, compared to other metrics, heart rate seemed to have a long latency before changes induced by TORs and such changes lasted for a long time as shown in Figure \ref{fig:HRPTimeseries}. This is consistent with previous studies as heart rate activities change gradually and required a longer time window to be stable \citep{solovey2014classifying,alrefaie2019heart}.


In summary, drivers' psychophysiological response patterns in the time domain are rather different to the same TORs. Some responded immediately and recovered soon while others had a long latency for responses and lasted for a long time. 
When we used the whole takeover transition period as the time window to calculate various measures for statistical analysis, it gave us an overview of drivers' states during takeover transitions. In contrast, analyzing the second-by-second time-series data gave us insights into their temporal changes and provided us recommendations on the optimal time window selection to improve the sensitivity and specificity of different psychophysiological measures.


\subsection{4.4. Limitations and future work}
First, to interpret the psychophysiological data, we compared our results with well-established literature and provided insights on drivers' cognitive load, attention, and emotion states reflected by psychophysiological data throughout the takeover transitions. Correlation analysis between drivers' dimensional emotions, subjective takeover performance, and physiological data was also conducted to increase the validity and interpretability of results. 
Future study can collect more self-reported measures on internal states (e.g., situational awareness) to help interpret the results.
It would also be valuable to examine the relationship between psychophysiological data and driving behaviors (e.g., minimum time to collision) to see whether psychophysiological data can be used to predict objective takeover performance. 

Second, given the fact that drivers' internal states are associated with multiple psychophysiological measures, we used several of them to reliably measure subtle changes in drivers' cognitive load, attention, emotional states, and situational awareness. However, a variety of psychophysiological measures can be derived from the raw physiological signals. For example, in addition to emotional valence and engagement, emotional arousal can also be predicted from facial expressions using machine learning  algorithms \citep{zhou2020fine}. In future studies, more metrics, such as emotional arousal, frequency-domain HRV, and fixations can be potentially included.


Third, we used a high-fidelity fixed-base driving simulator to imitate takeover situations in a controlled laboratory and recruited younger adults as participants. 
This is especially important when the psychophysiological measures collected are sensitive to various factors. However, the obtained results might be less ecologically valid than those obtained from on-road scenarios and across age groups. Future studies can replicate the experimental settings with naturalistic driving and recruit diverse drivers to see the robustness of psychophysiological measures. 

\subsection{4.5. Implications}

Psychophysiological measures indicated proactive responses induced by different NDRTs, traffic density and TOR lead time before performance behavior was observed. As a summary, the inclusion of psychophysiological measures helped provide insights into the often unconscious mechanisms underlying the takeover performance behaviors. Therefore, such measures can help researchers understand the mechanisms of takeover transitions by complementing other vehicle-related measures and improve predictions of takeover performance proactively. 


The reliable and valid assessment of drivers' internal states using psychophysiological measures can be the ground work to develop state detection and monitoring systems. Studies have shown that there are medium to strong  associations between psychophysiological measures and drivers' states \citep{du2020predicting,zhou2020driver}. Data from wearable devices can be used to train advanced machine learning models to indicate drivers' states in a continuous, non-obtrusive, proactive, and real-time way. Furthermore, according to monitoring results, an adaptive in-vehicle alert system can be designed to trigger warning or intervene drivers when sub-optimal internal states are associated with potential hazards during the takeover transition period.


\section{5. CONCLUSION}

This exploratory study systematically investigated drivers' psychophysiological responses to TORs in different NDRTs, traffic density, and TOR lead time conditions. During automated driving stage, we found that drivers had lower heart rate variability, narrower horizontal gaze dispersion, and shorter eyes-on-road time when they were in high cognitive load triggered by 2-back memory task. Upon the TOR, 4s lead time led to inhibited blink numbers and larger maximum and mean GSR phasic activation, indicating higher emotional arousal and stress than 7s lead time. Meanwhile, heavy traffic density resulted in significantly frequent HR acceleration patterns than light traffic density, suggesting ignorance of overwhelmed traffic information. 

While driving behaviors alone give us insights into drivers’ takeover performance, psychophysiological signals collected by non-invasive sensors allow us to estimate drivers' workload, emotions, attention, and situational awareness in a continuous and real-time manner. The findings provide us a broad picture of driver states throughout the whole takeover process and inform the development of driver monitoring system and design of in-vehicle alert systems in SAE Level 3 automated driving.

\bibliography{HFES-bibliography}





\end{document}